\documentclass[english,aps,prstper,reprint,showpacs,titlepage,longbibliography,floatfix,nofootinbib,superscriptaddress]{revtex4-2}   

\usepackage[T1]{fontenc}	
\usepackage[latin9]{inputenc}	
\usepackage{geometry}		
\usepackage{graphicx}
\usepackage{tabularx}
\usepackage{subfig}
\usepackage{amssymb}
\usepackage[above,below]{placeins}	
\usepackage{times}

\usepackage{placeins}
\usepackage{caption}
\usepackage{makecell}

\usepackage{hyperref}  
\hypersetup{colorlinks=true,urlcolor=blue,citecolor=blue,linkcolor=blue}   
\usepackage{enumitem}          
\setlist{nosep}                 
\begin{document}

\begin{titlepage}

  \title{Studying Impact and Intent of Design: Conjecture Mapping for Affect-Centered Analysis}

\author{Sarah McHale}
    \affiliation{Department of Engineering \& Physics, Providence College, 1 Cunningham Sq, Providence, RI, 02908} 
    \affiliation{School of Physics and Astronomy, University of Minnesota -- Twin Cities, 116 Church St. SE, Minneapolis, MN, 55455} 
  \author{Tor Ole B. Odden}
    \affiliation{Department of Physics, Center for Computing in Science Education, University of Oslo, 0316, Oslo, Norway} 
  \author{Ken Heller}
    \affiliation{School of Physics and Astronomy, University of Minnesota -- Twin Cities, 116 Church St. SE, Minneapolis, MN, 55455} 


  \begin{abstract}
Physics education researchers have argued that authentic physics education includes computation as part of a physics student's training, and many parties have made efforts towards this goal. However, most research on this teaching modality has centered cognitive impacts rather than affective impacts, so little is known about the affective outcomes of holistically integrating computation into physics courses. To address that need, we present a case study of a multi-day activity within a computationally integrated modern physics laboratory course. With data from ethnographic course observations and semi-structured interviews, we introduce a novel modification of conjecture mapping to distinguish between the professor's intent behind the activity design and the impact on a student's physics computational literacy and physics identity. 
In doing so, we highlight the methodological suitability of conjecture mapping for comparing the intent and impact of curricular design and explicate the specific misalignments that led to different affective outcomes than intended. 
    \end{abstract}

  \maketitle
\end{titlepage}

\section{Introduction}\label{Sec:introduction}
For years, physics education researchers and physics organizations have argued that authentic physics instruction includes training in scientific computation~\cite{caballero2020advancing, caballero2024computing,behringerengelhardt, behringer, oddencaballero, landau, chonackywinch}. Numerous individual instructors and institutions have taken steps towards this educational goal in various ways (e.g.,~\cite{roundy2015look, fuller2006numerical, landau, chonackywinch, gavrin, picup, roos, paradigms, caballero2014model, StThomas, cook2008computation, serbanescu2011putting}). 
One such approach is to integrate computation into a sequence of physics courses so that students encounter computation in a physics context, rather than as a stand-alone course in computer science. In this model, computation is deeply and deliberately scaffolded across the curriculum so that students gain both computational skills and physics knowledge simultaneously. 

Recent scholarship has investigated the outcomes of these kinds of curricular changes. Most of this literature either focuses on describing particular curricular approaches~\cite{roos, serbanescu2011putting, paradigms, caballero2014model, StThomas, cook2008computation, odden2020using, Irving_2017, caballero2014integrating, orban2018hybrid, atherton2023resource, oddencaballero, tufino2025using} or analyzing the cognitive effects of this teaching modality 
\cite{araujo2008physics, CTframework, mackessy2021comparing, weatherford2013student, oddenzwickl, phillips2023physicality, lunk2012framework, kohlmyer2005student, oleynik2019scientific}. 
However, there are other important dimensions of physics learning that remain understudied in a computational context. 

In both computer science and physics, scholarship has investigated the cognitive impacts of 
emotions~\cite{haussler2000curricular, lishinski2017students, gupta2018exploring, bodin2012role}. However, affective experiences are a salient feature that students remember about their encounters with computation, and students sometimes retrospectively judge themselves poorly even if they experienced success~\cite{kinnunen2011cs}. Further, the ways in which students internalize definitions of success shape their identity formation~\cite{bumler2019previous}, so helping students have positive and meaningful experiences with computation necessitates more than fostering success. As students learn about physics, they also learn about themselves as physicists and develop affective stances towards themselves relative to their field of study. While existing scholarship has operationalized affect in computationally integrated physics courses through self-efficacy and attitudes towards computation~\cite{tufino2024integrating,gallagher2026evaluating}, it has not described how students develop their identity in relation to computational tasks. We must understand this development in order to characterize uses of computation that effectively and equitably meet instructional goals~\cite{caballero2020advancing}; we argue that the development of a positive self-perception is an instructional goal, even if latent.

At present, few studies have examined this relationship. 
To investigate the development of these affective stances, Hamerski, et al. studied challenges from the perspectives of high-school students in a computationally integrated physics course~\cite{hamerski}. Interviews with and observations of these students revealed that computation is a double-edged sword, with the potential both to provide authentic learning for students with preexisting familiarity and to bring about an aversion to physics for those without. The importance of prior computational exposure also surfaced in Bumler, et. al's study of college students in a computationally integrated introductory electricity and magnetism course as they developed new academic identities and redefined success in this setting~\cite{bumler2019previous}. Whereas both of these studies 
foreground student experiences, we aim to understand when, how, and why instructor and student perspectives differ to further align intent and impact so we can provide insight on design decisions that leverage computation as identity-affirming.

However, for instructors, it is both challenging and complex to design components of a computationally integrated physics course so that students have positive affective experiences because the ways in which instructors set out to impact their students do not always come to fruition. A computationally integrated course consists of pieces of a whole that are meaningfully tied together. As such, it is important to gain a qualitative understanding of what granular features of encounters with computation help students have an affirming experience in ways that instructors can implement. 
We therefore need to learn how the design of computational activities affectively shapes the student experience, distinguishing the intent behind instructors' decisions from the impact on students. 



The case study explored in this paper is part of a broader study that investigates the affective impacts of integrating computation throughout the physics major at the University of Minnesota -- Twin Cities, through mixed-methods surveys, semi-structured interviews, and ethnographic observations~\cite{mchale2024students}. During exploratory analyses, we identified a noteworthy theme: students' experiences often differed from the instructors' intentions. In the present study, we zoom in on the first course involved this change effort and specifically focus on a multi-day activity that embodied the aforementioned theme. To gain a suitable depth of information on the complexity of affective experiences, we center this case study on two individuals: the professor who designed the activity and a student who tied a shift in in her self-perception to engagement with this activity. 
Therefore, this paper aims to address the following research question: \textit{How does the affective intent behind the design of a computational activity differ from its impact on an individual student's experience?}

To address our research question, we modify the analytic method of conjecture mapping~\cite{sandoval} to evaluate when, how, and why the instructor's intent behind activity design differed from a student's affective experience during a multi-day activity. 
We begin by explicating how the professor concretized her intentions by representing design features, expected mediating processes, and outcomes of student self-perception through physics computational literacy and physics identity. We then create linkages between categories in the form of design and process conjectures that represent the professor's expectations for student engagement and self-perception. Then, using the same categories, we analyze a student's perspective by mapping how her experience affirms, negates, and complicates the professor's design and process conjectures. As such, this modified use of conjecture mapping enables us to tie the complexity of affective experiences to an activity design in a rigorous and testable manner, thereby providing insight to both design and theory.


\section{Theoretical Frameworks}
\label{theory1}
This study takes an affect-based approach to investigate the role that a computationally integrated curriculum plays in the potential for a multi-day activity to serve as an affirming experience for students. In particular, we are interested in understanding how integrating computation into the teaching of physics may shape students' self-perception relative to the field of physics. We therefore use two theoretical frameworks to operationalize affective outcomes: physics computational literacy and physics identity.
\subsection{Physics Computational Literacy}
In \textit{Changing Minds: Computers, Learning, and Literacy}, education researcher Andrea diSessa claimed that computation is a new form of literacy, of the same importance as reading, writing, and math~\cite{disessa_2000}. In so doing, diSessa created the term computational literacy (CL) to represent the idea that computation enables new forms of learning and understanding. This framework posits that to develop a robust CL, three components must be tended to:
\begin{itemize}
  \item \textbf{Material CL:} familiarity and fluency with computing;
  \item \textbf{Cognitive CL:} applying computing to think differently; and
  \item \textbf{Social CL:} communicating with or about computing.
\end{itemize}

Odden, Lockwood, and Caballero adapted CL in the context of physics, coining the term physics computational literacy (PCL) \cite{odden2019pcl}. Their study focused on how a teaching tool proposed by diSessa, computational essays, demonstrates the suitability of the framework to disentangle the three deeply interconnected pillars to both characterize educational approaches and explicate student difficulties in a computationally integrated physics course. Applying diSessa's three pillars of CL, they categorized physics-specific examples from their dataset including, but not limited to:
\begin{itemize}
  \item \textbf{Material PCL:} familiarity with syntax, debugging code, algorithmic thinking, preferences for certain programming languages, beliefs about structure of code, etc.;
  \item \textbf{Cognitive PCL:} modeling practices, extracting physical insights from code, knowledge of suitability of computational methods, beliefs about how computation can be used in physics, etc.; and
  \item \textbf{Social PCL:} organizing code, commenting code, use of computational notebooks, beliefs about presentation of computational analyses.
\end{itemize}
In particular, they highlighted that PCL exists and manifests at a variety of levels: practices, knowledge, and beliefs. Defined as attitudes or value judgments about computation, PCL beliefs specifically tend to affective outcomes by influencing the uptake of practices and development of knowledge~\cite{odden2019pcl}.

To test diSessa's claim that a literacy needs a literature, Odden and Zwickl used conjecture mapping to demonstrate how computational essays aid students in developing a robust PCL through engagement with and production of student-generated computational literature~\cite{oddenzwickl}. 
This study primarily centered students' concrete interactions with the construction of code, thereby focusing on PCL practices and knowledge, leaving the development of PCL beliefs understudied. This trend extends across much of the existing PCL literature~\cite{odden2019pcl, oddencaballero, fredly2026physics, cammarota2025social}, although Nearhood and Hamerski have begun to explore how beliefs shape social PCL practices~\cite{nearhood_student_2025}.

To highlight affective impacts of integrating computation into a physics course, we focus on how specific design decisions impact the development of PCL beliefs -- how students feel about their own PCL and how PCL is supported within the culture of their computationally integrated physics course.

\subsection{Physics Identity}
We also consider the extent to which students see themselves as physicists or `physics people' as an affective outcome. We operationalize this through the construct of physics identity (PI)~\cite{hazari2010, hazari2020}.
This framework originates from a study of successful women of color in science, in which Carlone and Johnson developed the framework of science identity \cite{carlonejohnson}. They found three factors meaningful for developing and sustaining a science identity: performance, competence, and recognition. To specify this framework for physics students, Hazari, et al. introduced interest as another factor and quantitatively revealed the co-existence of performance and competence as one factor~\cite{hazari2010}. When applying this framework to women in various stages of their physics majors, Hazari, et al. later found that sense of belonging differentially impacted students' physics identities \cite{hazari2020}. In this study, we therefore consider PI through the following components: 
\begin{itemize}
  \item \textbf{Performance/Competence:} belief in one's ability to understand physics content and/or to perform physics tasks;
  \item \textbf{Recognition:} being recognized by others as a physics person / one's perception of how others view them in relation to physics;
  \item \textbf{Interest:} personal desire or curiosity to learn and/or understand more physics; and
  \item \textbf{Sense of Belonging:} one's perception of fitting in or not feeling excluded in the/their physics community.
\end{itemize}

While PI has been studied in a variety of physics contexts~\cite{quichocho, kalender2019gendered, hyaterCPI, hazariCUWIP, wangHSPI, mcdermott, fracchiolla2020community}, it has rarely been studied in computationally integrated physics courses. Addressing some of the same constructs, Lane \& Headley conducted semi-structured, retrospective interviews with five undergraduate students in a variety of upper-level courses that used computation in homework assignments \cite{LaneHeadley}. Through a Communities of Practice (COP) framework, they studied how physics students develop expert-like perceptions of computation 
and found that students struggled to learn a second coding language, to feel confident developing code from scratch without help from their peers or the internet, and to engage when they did not view assigned tasks as important to physics. Many of their findings relate to competence as a central pull towards identity formation through modes of belonging, in alignment with the COP framework~\cite{PhysRevPhysEducRes.16.020143}. This may suggest that other components of physics identity can play important roles in the development of positive self-perception when studied through alternative lenses. 

Studying the development of students' PI in a computationally integrated physics course beyond the introductory sequence provides the opportunity to ascertain factors that positively or negatively impact PI not only in a new context that allows for students to redefine physics, but also in a manner that foregrounds the continued evolution of a physics identity. Integrating computation into the physics curriculum may change students' notions of what it means to be a physicist and the extent to which they identify with the field. 
So, in this study, we aim to understand affective impacts of integrating computation into a physics course by examining the role that the design of a multi-day activity plays in the development of students' physics identities. 

\section{Methods}\label{Sec:methods}

\subsection{Data Sources}
For studies that investigate affect, as this one does, we rely on data sources that can provide suitable depth and richness to understand how individuals experience the forces underlying their emotions. To center participants' reflections on their own experiences, we prioritize data collected from semi-structured interviews with students and instructors of this computationally integrated physics course. To increase the credibility of our interview analyses, we triangulate our data collection methods to include mixed-methods surveys, ethnographic course observations, and document analysis.

As part of a broader departmental effort to integrate computation across the physics major, the following surveys were administered to students at the beginning of the semester: a demographic survey, the CLASS \cite{class}, the SOSESC-P survey \cite{sosescp}, and a mixed-methods survey asking about students' experiences with and beliefs about computation. All surveys were administered at the beginning and end of the semester with a nearly 100\% response rate. Students were given points towards their overall homework grade for completion, but only the first author saw students' responses and scores. To identify representative candidates for further study, the first author analyzed these survey responses and invited students to interview who comprised a range of demographics, learning attitudes, self-efficacy, and viewpoints on and experiences with computation. Five students from this course agreed to participate in a two-part interview series, one interview midway through the semester and one towards the end. The first and third authors crafted semi-structured interview questions to elicit feedback on students' holistic experience in the class and on their affective experiences related to the constructs of PCL and PI. The first semi-structured interview was conducted early in the semester and concentrated on students' expectations for the course, previous experiences with computation before the class, and their foundational self-perception. The end-of-semester interview adjusted focus to students' reflections on their experience in the course and, in their opinion, resulting impacts on their PCL and PI. Student interviews lasted approximately 1-1.5 hours, and recordings were transcribed and deidentified for analysis.  

The professor and both TAs of this course were also invited to and agreed to participate in a similar 2-part interview series. Again, the first and third authors crafted semi-structured interview questions to elicit instructors' perceptions of the course, their design decisions, and the role they believed those both played in their students' experiences. To characterize instructors' perspectives coming into the course, the semi-structured interview protocol for the initial interview asked instructors about the course culture, their interpretation of class goals, their assessment of the course evolution through the change process, their personal software experience, and their demographics. The semi-structured interview protocol for the end of the semester asked  instructors about their reflection on the success of the semester, pertaining to both the culture of the course and the integration of computation into the course, as well as their rationale behind instructional decisions and reflections on their students' PCL and PI. Instructor interviews lasted approximately 1-1.5 hours, and recordings were transcribed and deidentified for analysis.

To gain insight into the embodiment of the student and instructor perspectives, the first author also conducted ethnographic course observations of the labs, lectures, and TA meetings associated with this course and recorded fieldnotes while observing~\cite{emerson}. As a former TA of the course, the first author interacted with students and instructors during these observations when questions arose. This built rapport with the students, who sometimes shared their thoughts on the course with the first author during observations. In total, the raw, ethnographic data collected amounted to approximately 33 hours of observation and 46 pages of fieldnotes.

Beyond conducting classroom observations, the first author also had access to the course Canvas page and analyzed materials like class announcements, homework assignments, and lecture slides. These document analyses served to round out our characterization of the multi-day activity that is the subject of this paper and triangulate participants' explanations of their associated expectations~\cite{bowen}. 

During both the data collection and analysis stages of the project, 
the first author used a researcher journal to reflexively understand and account for her role in the research process~\cite{ortlipp}. Additionally, the first author also wrote analytic memos throughout analysis to tease out the "fundamental stories in the data" \cite{tracy, SaldanaJohnny2016Tcmf}. In so doing, the first author noticed an emergent theme: students' experiences often differed from instructors' intentions in complex ways. 

During a preliminary thematic analysis of the end-of-semester interviews, the first author noted that by and large, students felt this computationally integrated physics course was coherent. However, one student specifically mentioned a multi-day activity as a salient experience when asked about the development of her PCL and PI, with a different affective tone than she spoke about the course as a whole. In the form of a comparative case study, this paper analyzes that student's experience with the multi-day activity and specifically centers her conceptualization of the coherence between this activity and the course as a whole. We contextualize her perspective with that of the professor to elucidate how and why the design of this activity manifested differently than intended. The professor and student have been deidentified with pseudonyms, Professor Evans and Bridget, respectively. Per the participants' requests, we use gendered pronouns for them throughout this paper.

\subsection{Conjecture mapping}
The goal of this study is to distinguish between the intent and impact of the design of a single multi-day activity. As such, analysis calls for a method that can track continuities and discontinuities from design decisions to outcomes. 
To do so, we take inspiration from design-based education research and use the analytic method of conjecture mapping, which explicates salient features of a learning community and outlines how they function together to generate outcomes \cite{sandoval}.
\subsubsection{Components of a conjecture map}

\begin{figure*}
{\includegraphics[width=\linewidth]{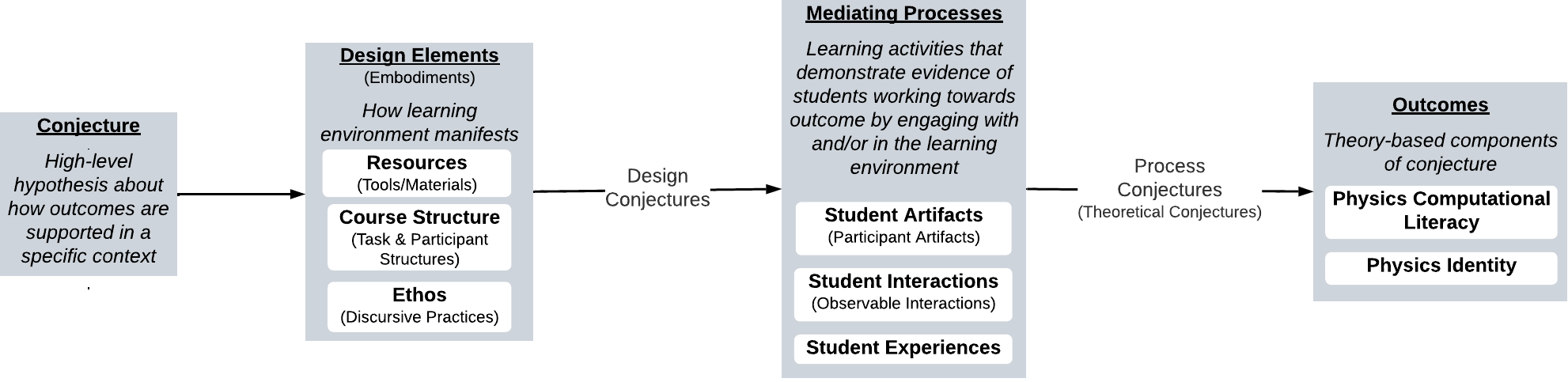}}
  \caption[Components of a conjecture map, as used to evaluate design and process conjectures.]{Components of a conjecture map. When we use a modified term, Sandoval's original term is included in parentheses~\cite{sandoval}.} 
  \label{fig-ConjectureMap}
\end{figure*}

Conjecture mapping begins with a high-level hypothesis about how outcomes are supported in a specific context, known as the conjecture. This conjecture connects design elements that embody how the learning environment manifests. These embodiments of the design are taken up by students through mediating processes that demonstrate engagement in and with the learning environment. These mediating processes also provide evidence of students working towards the outcomes and on a conjecture map, they directly connect to the outcomes that align with the theoretical components of the conjecture. 

In his original proposal of this method, Sandoval connected components of a conjecture map through design conjectures when connecting from an embodiment to a mediating process and through theoretical conjectures when connecting from a mediating process to an outcome. In so doing, he highlighted the iterative nature of conjecture mapping, as these conjectures were designed to produce refinements both of design and theory. Sandoval considered this dual power to be particularly useful for design-based research. However, we use the method retrospectively, and as such, our application of the method differs slightly.

To more tightly tie our conjecture maps to this specific activity under study, we renamed some of Sandoval's original terminology. However, the meaning and role of these categories remains the same. Fig.~\ref{fig-ConjectureMap} details how we use a conjecture map in this study and maintains Sandoval's terminology in parentheses.

We refer to embodiments as design elements to center the role of activity design in the development of affective outcomes, analogous to the modification used by Boelens, et al~\cite{boelens2020conjecture}. Within the design, we considered the structure of the course, the resources intended as support, and the underlying ethos. These renamed facets span the same theoretical space as Sandoval's subcategories of tools/materials, task \& participant structures, and discursive practices.

Similar to Kim and Saplan, we reiterate the centrality of students by renaming participant artifacts to student artifacts~\cite{kim2024making}. Because students also worked on portions of this activity outside of class, and thereby outside of observation periods, we broadened the category of observable interactions to student interactions, thereby including interactions that were observed by the first author or reported by participants. However, these two categories did not encompass all of the means through which students enacted the professor's expectations for the activity, so we added the subcategory of student experiences to account for experiences that were not strictly artifact- or interaction-based~\cite{boelens2020conjecture, kim2024making}.

In this paper, we propose conjecture mapping as a suitable analytic tool to distinguish between instructor intent and student impact. 
Rather than continually revising the same conjecture map throughout the re-design of an activity, we create a conjecture map for each narrative that describes a different perspective on the same multi-day activity and analyze how and why the paths towards the outcomes differ. 

We want to understand the aforementioned paths in two ways: why mediating processes are used with engagement through the learning environment, which is a design conjecture; why the mediating processes lead to outcomes, which more closely aligns with McKenney and Reeves' reframing of theoretical conjectures as process conjectures~\cite{mckenney2018conducting}. So, in adopting this terminology, we therefore compare how the professor intended the design and process conjectures to be enacted with the student's enactment of the design and process conjectures. The resulting visual representations help clarify how the use of certain design elements led to the use of different mediating processes, which in turn fostered the development of specific outcomes. 

We additionally nuance our evaluation of design and process conjectures by tracking whether the connection has a positive or negative underlying tone, known as valence~\cite{russell1999bipolarity}. While this distinction simplifies complex emotional experiences into a binary, it enables us to represent actions that individuals take or beliefs that individuals develop while feeling negatively about their decision to do so. We also use this to represent actions or beliefs that individuals struggle with, again differentiating between the positive affective experience of eustress or a negative affective experience of distress. This representational distinction is a departure from the original form of conjecture mapping, in which the design elements and mediating processes themselves would change to describe engagement. Including valence enables us to test design and process conjectures in a manner that accounts for hedonic variation in the student's enactment of the professor's design and process conjectures. Further, when a participant expresses both positive and negative affect, we choose to represent both so that we do not allow positive and negative valence to cancel out, thereby misrepresenting neutrality~\cite{Varela2005-VARATS}.

\subsubsection{Creating and evaluating a conjecture map}
As a research team, we found it important to bring together our varied data sources through an analysis that was rigorous and credible at all stages~\cite{tracy2010qualitative}, from the initial coding to the eventual creation of our conjecture maps. A quantitative approach like inter-rater reliability would not function well with the interpretive process of creating conjecture maps, so we instead used a consensus-based approach to the interpretive process of intercoder agreement ~\cite{tracy,SaldanaJohnny2016Tcmf}. By iterating until our understanding of the data converged to a shared understanding, our discussion-based methodology maintained the rigorous spirit of the commonly-used quantitative approach of inter-rater reliability~\cite{o2020intercoder}.

To trace the development of affective outcomes, we first explicated how the individuals under study experienced the activity. 
In alignment with the study's consideration of these outcomes through two theoretical frameworks, the first author coded all interview transcripts deductively using the subconstructs of PCL and PI addressed in Section \ref{theory1}. The first author then selected portions of the interviews that pertained to the activity under analysis and distributed the relevant data to the third author, who then independently coded the data while referring to a shared codebook. The first and third authors compared independent codes, discussing until convergence. These deductive codes served as the outcomes for all of our conjecture maps, as they address the theoretical components of the initial conjecture.

Because Professor Evans controlled the design of the multi-day activity, we used her interview as the primary data source to populate subcategories of design elements and mediating processes. We also used observational data and analysis of course documents to both triangulate Professor Evans' interview and provide a more robust description of the activity design. Beginning with the professor's data, the first author used the reliable deductive codes as outcomes on the conjecture map and connected the content of the quotes from outcomes to the actions Professor Evans expected students to take, which we categorized as mediating processes. These connections thereby created theoretical conjectures in a backwards design. To connect the mediating processes to design elements, the first author again initially referred to interview data to prioritize concrete explications of design rationale. Such evidence was not always present, as these semi-structured interviews covered participants' experiences in the course in general, so the first author also used data from observations and document analysis. 

To ensure that these design elements and mediating processes were named in a credible and representative manner, the first author recorded rationale for categorization in analytic memos and iterated over terminology with the research team until consensus. We used these same terms to populate the student's conjecture map, and using the same components for each conjecture map allowed our analysis to center the different manifestations of design and theoretical conjectures on each map. As such, our analysis aligned with our goal to investigate the impacts of intended or actualized engagement with design.

While making the design and process conjectures from the professor's data, the first author collapsed connections into themes, which form the foundation of the textual description of the design and process conjectures. To rigorously and robustly evaluate the manifestation of the conjectures, the first author used these themes to independently trace the student's enactment of each respective design and process conjecture. So, the design and process conjectures visualize the similarities or differences between intent and impact by depicting each participant's distinct linkages representing the same theme. 

As with the creation of the professor's conjecture map, the creation of each connection relied heavily on interview data to prioritize the student's explicit reflection on her own experience, and data from observations and document analysis were supplemented as needed. Across analysis for both participants, the first author logged rationale for these decisions in analytic memos. 

As is typical in the creation of conjecture maps, this was an iterative process (e.g.,~\cite{kim2024making}). All three authors discussed the suitability both of the textual description of the conjectures and the linkages serving as the visual representation. Creating the design and process conjectures was a consensus-based practice, in which all authors studied the data and characterized the affective connection as either positive or negative. At this stage, the authors discussed each conjecture map individually, rather than comparing the two maps immediately. 

All linkages were included in initial iterations of the conjecture maps, and data sources were connected to each linkage and mediating process, so initial maps were highly complex and included all data relevant to each individual's experience with the activity. Through conversations across the research team, the authors came to consensus on the most salient themes on each conjecture map and correspondingly simplified the conjecture maps to only display those connections. In this paper, we only present elements of the conjecture maps that relate to an emergent theme at the heart of our conjecture: the ways in which the multi-day activity related to the coherence of the course, from the point of view of both the student and the instructor.

With rigorous and robust conjecture maps in hand, the first author compared the design and theoretical conjectures by focusing on how they aligned or diverged across the two conjecture maps. Again, these analyses were shared with the second and third authors, who discussed until reaching consensus on the implications of the different linkages for design and process conjectures.

\section{Comparative Case Study}\label{Sec:case_study}
To elucidate the ways in which affective intent and impact can differ in encounters with computation in physics, we situate our comparative case study~\cite{alma9975672698201701} in a computationally integrated modern physics lab course. In specific, we investigate a student's affective experience with a multi-day activity and contextualize it with the professor's intent for the activity. As we are interested in how the activity connected to the course as a whole, we first characterize the broader learning environment in which this activity occurred.

\subsection{Context}
The School of Physics and Astronomy at the University of Minnesota -- Twin Cities has committed to integrate computation throughout the physics major, with the exception of large-enrollment introductory courses dominated by students from other disciplines. As part of this change process, a computation committee was formed to help instructors map the learning goals of upper-level courses and assess if computational material aids in the achievement of these learning goals. In particular, this paper focuses on the first course involved in this change process, a sophomore-level modern physics laboratory  course, at a time in which it had already undergone the aforementioned review process. Situating our study in a context enables us to investigate a curriculum that has already been revised to suitably integrate computation and is in the process of fine-tuning, thereby minimizing the potential for the novelty of computation to overwhelm other findings.

Each week, the professor led a a 50-minute lecture, with a 39:1 student:teacher ratio. Students sat at tables arranged in long rows parallel to the whiteboard and projector screen at the front of the room. Throughout the semester, Professor Evans regularly began class with a series of dense slides that covered the statistical methods students were to use in the week's homework and labs. After deriving the relevant equations, she used data from students' lived experiences -- like baseball batting averages, the rates at which fireflies flash, or experiments students had conducted in lab -- or from her own research as sample datasets to contextualize her calculations and conclusions. As Professor Evans flew through her slides, students were either hunched over frantically writing copious notes or reclined back in their chairs absorbing material with no writing utensils in hand.

Professor Evans' lectures often included sample problems that she worked through on the whiteboard or through animated slides. Occasionally, she would pause to let students conduct calculations and when she did so, she always verbally encouraged students to turn to their neighbors and work together. Even though students chose their own seating and chatted with each other before the start of class, they rarely responded to Professor Evans' requests for collaboration, and the class often remained as silent during group work as it was during lecture, only turning to a quiet murmur of conversation after repeated prodding from Professor Evans. Every now and then, the course's Teaching Assistants (TAs) sat in the back of the room to observe lecture, but they remained seated and did not circulate the room as Professor Evans did during in-class activities. After the end of almost every lecture session, students approached Professor Evans to ask questions about their homework, her lecture, or their lab analyses, and Professor Evans always responded genially to every student until the queue of students left satisfied.

This course also had a weekly lab and a weekly TA meeting. Every week, each TA led a 3-hour lab 
with a 13:1 student:TA ratio. In the first week of lab, students completed scaffolded workshops specific to the computational platform of the course, MATLAB. During the second week of the course, students worked in pairs to conduct an experiment common to all students to practice using MATLAB in an experimental context. For the remainder of the semester, students worked individually or in pairs to complete 5 2-week labs. Attendees of the TA meeting included more than just the course professor and TAs, as a member of the department's computation committee joined in on the discussions of student challenges, achievement of educational goals, and plans for the upcoming week.

Considering the broader dataset of interviews with students in this course -- not only the focus of this case study, Bridget -- participants often referenced the fast-paced nature of the course and the difficulty juggling statistical, computational, and physics knowledge across components of the course. However, many concerns were assuaged by the coherence of the course, as students felt that what they learned in one component of the course, like lecture, benefited them in another, like lab. Students expressed that this helped them feel at ease with the broad coherence of the course, and they felt supported both by Professor Evans and by the resources she provided, like posted lecture slides. However, one participant, Bridget, spoke about both the course and herself with a markedly different affective tone in her end-of-semester interview than in her early-semester interview. Capitalizing on the flexibility of a semi-structured interview, the first author asked Bridget about the development of these feelings, and Bridget continually mentioned how she felt bad about herself 
during and after a single multi-day activity, even though she liked the course as a whole and found the course more broadly to be affirming. Bridget was the only interviewee for whom this activity seemed to shape her perception both of herself and of the course, and given the goals of the broader project in which this study occurred, we decided to analyze her experience to understand why she developed a negative affective stance.

The multi-day activity to which Bridget referred occurred towards the end of the semester. Professor Evans 
structured it as two homework assignments, with an in-class workshop in between. In the first associated homework assignment, students expanded upon an earlier homework assignment in which they linearized a power law relationship and determined fit parameters -- slope and intercept -- with uncertainties. In this new assignment, students first calculated $\chi_{r}^2$ from fit parameters and uncertainties they found on the earlier assignment. Then, to more systematically investigate how $\chi_{r}^2$ changed with deviations from the best fit, students computationally calculated $\chi_{r}^2$ for a power law by looping over one fit parameter while holding the other fit parameter constant, thereby not requiring linearization. Students produced 2-dimensional plots of $\chi_{r}^2$ vs. each varying fit parameter. Then, they expanded the same method to three dimensions and used nested \texttt{for} loops to vary both fit parameters over their respective range of uncertainty from the prior homework. Students also plotted a 3-dimensional $\chi_{r}^2$ surface and explained what the graph communicated. 

While students had linearized data throughout the semester, they had not used \texttt{for} loops in any capacity. However, in lecture the week before this assignment, Professor Evans directed students towards online references to learn about \texttt{for} loops, both from the class and the internet.

The in-class workshop asked students to annotate code written by one of the course's TAs that conducted a 2-dimensional grid search to find the fit parameters that minimized $\chi_{r}^2$. Students' annotations were supposed to both explain how each section of the code worked and bracket each of the for loops to "match ends." Students were also asked to determine what changes the program needed to fit to the error function. The first author observed this course session and noticed that Bridget's engagement was similar to most other students.

These expectations for the in-class workshop were also components of the second associated homework assignment. In addition, the second assignment asked students to once again create a 3-dimensional plot of the $\chi_{r}^2$ surface and overlay the path taken by the grid search. This assignment was the last homework of the semester.

\subsection{Intent: Professor Evans' Perspective} \label{intent}
Professor Evans is a physics professor who conducts research in experimental particle physics. She had taught this course multiple times before the semester under study and was regularly involved in the improvement of this course, even before the concerted effort to integrate computation into the curriculum. Both generally in this course and specifically in this multi-day activity, Professor Evans wanted her students to learn 
computation and feel like physicists. As with many other components of this course, she revised this activity from how she had used it in previous years. After a focused effort on improving the activity for student outcomes, Professor Evans had high hopes for the activity. 
She viewed this as a capstone that extended what students had learned thus far in class, and she believed that bridging course content and real-world physics empowered students.

\begin{figure*} 
  \includegraphics[width=\textwidth]{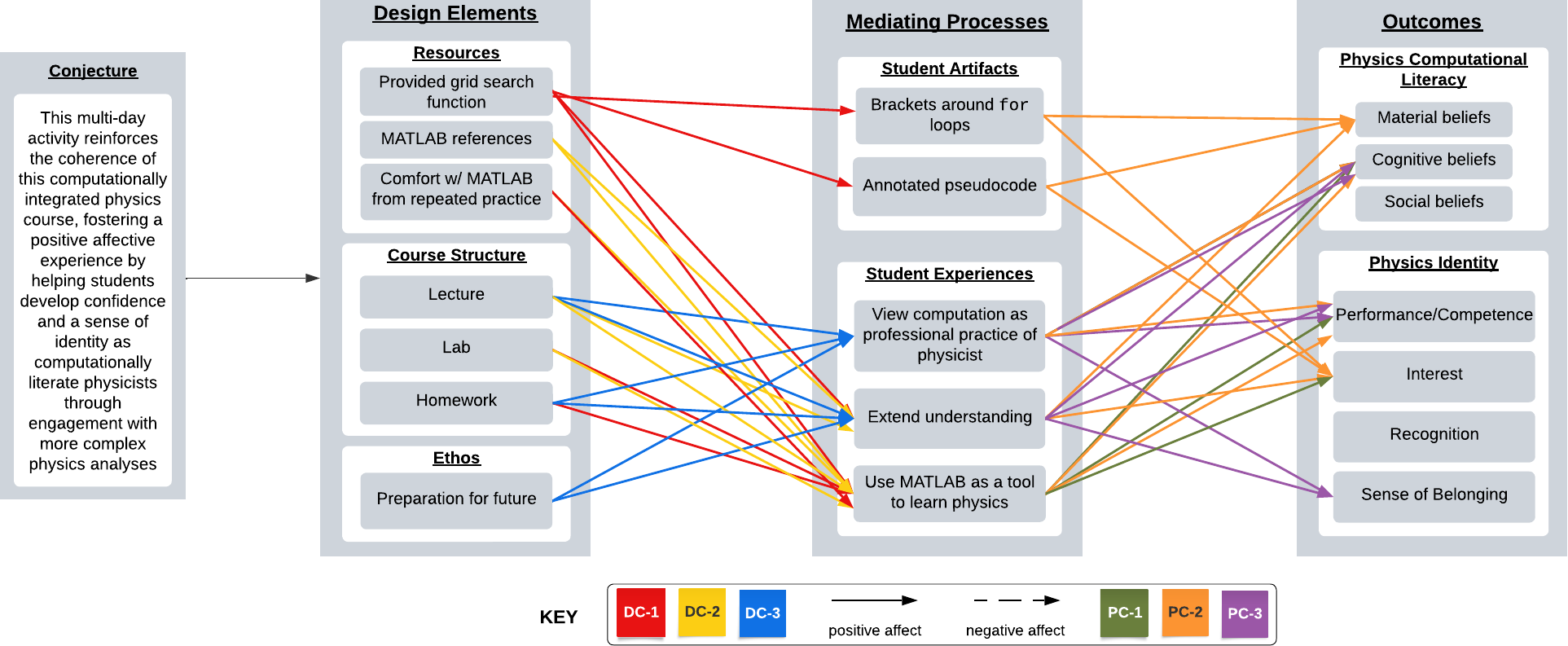}
  \caption[Professor Evans' conjecture map, representing intended coherence between the entire course and the activity under study]{Conjecture map representing Professor Evans' perspective on the multi-day activity, specifically depicting intended coherence between the entire course and the multi-day activity under study.}
  \label{fig-intent}
\end{figure*}

In what follows, we present the creation of a conjecture map representing Professor Evans' intentions for the multi-day activity, as shown in Fig.~\ref{fig-intent}. We begin this conjecture map by proposing a hypothesis that emerged from our preliminary analysis of Professor Evans' data: \textit{This multi-day activity reinforces the coherence of this computationally integrated physics course, fostering a positive affective experience by helping students develop confidence and a sense of identity as computationally literate physicists through engagement with more complex physics analyses.}  

The connections on our conjecture map detail how she reinforced the messaging underlying her connection of this activity both to the coherence of the course and to real-world physics. We first highlight how her design features link to mediating processes through design conjectures (DC) and characterize the underlying affective tone as positive or negative. To track how she intended to develop students' affective experiences, we use the constructs of our theoretical frameworks for affect, PCL and PI,  as our outcomes. We then trace Professor Evans' perspective through connections from mediating processes to outcomes in process conjectures (PC), again delineating between positive and negative affective tone. The specific details of the data sources are summarized in 
Appendix \ref{Sec:Evans_data}.

\subsubsection*{Design Conjecture 1: If computational tasks are appropriately scaffolded, then students will be able to use computation as a tool to learn physics.}
Professor Evans was newly involved with her department's effort to integrate computation into its physics curricula. As part of this effort, she deliberately integrated computation into this course with scaffolded computational activities so that students would gradually build up to writing more complex code. She had students initially use built-in MATLAB functions for linear fitting and later write their own linear fit functions. In so doing, she hoped students would "use the canned programs to begin with so they can just see how it works" and then "learn what the statistics is and then they can put the two together."

In our observations and document analysis, we saw structured and repeated use of MATLAB in the laboratory and homework. Some homework assignments explicitly stated to students that what they were asked to do was "directly
applicable to the Labs you will be doing this semester. Feel free to use the MATLAB code you develop here directly in the relevant lab analysis documentation." Every lab analysis required students to use MATLAB to calculate and report their results. Some experiments, like x-ray diffraction, even asked students to use different computational techniques depending on the week of the semester in which students conducted the experiment. When new computational approaches like linear fitting were introduced, students often first encountered them in the homework and then applied them in lab. Therefore, DC-1 connects from the design elements of \textit{Homework}, \textit{Lab}, and \textit{Comfort w/ MATLAB from repeated practice} to the mediating process of \textit{Use MATLAB as a tool to learn physics} with positive affect.

Professor Evans' familiarity with the course enabled her to revise course material so students would more effectively meet her educational goals. During a TA meeting, Professor Evans said she felt that versions of this activity from previous years only asked students to produce a $\chi_{r}^2$ surface rather than truly conduct a grid search, so she revised this activity to provide a grid search function for students to use, which she had one of the TAs write. This re-structuring scaffolded the workshop so students would not need to write this more advanced code from scratch and the entire multi-day activity would build on itself. However, this provided code used nested \texttt{for} loops, which the students had never used in class. As "an example of using someone else's code," Professor Evans expressly asked students to annotate the function to both bracket each individual for loop and explain each line of code. So, DC-1 also connects from the resource of \textit{Provided grid search function} to the experiences of \textit{Use MATLAB as a tool to learn physics} and \textit{Extend understanding}, as well as the artifacts of \textit{Brackets around for loops} and \textit{Annotated pseudocode}.

These dual manifestations of DC-1 highlight the professor's perceived coherence between the activity and the course.

\subsubsection*{Design Conjecture 2: If students learn statistics in lecture, they can apply them when writing code.}
The first author regularly observed Professor Evans' lectures and repeatedly saw her cover statistics at length. 
However, Professor Evans felt constrained by the structure of the course, with only 1 lecture per week. Because students wrote code for their lab analyses, she felt that it was important to use lecture to explain and demonstrate the statistics they would implement computationally:
\begin{quote}
    This is not a course on MATLAB. It is a course on how to do an experiment and the statistics [for analysis]. And I feel that with only one class a week, I can't split my time between that and also actually teaching them MATLAB. So one has to hope that this structure will tide them through where they learn it at the beginning, they have resources that they can go back to, and that they learn by doing over and over and over again.
\end{quote}

In this passage, Professor Evans references the way in which she restructured the lab component of this course. As was not the case in previous years, the first few weeks of lab served as an on-ramp to MATLAB, where students focused on learning syntax and how to present computational analyses. She also mentions MATLAB "resources" that she wants students to use, which our document analysis and observations connected to course-specific manuals and online help sites that she directed students towards both during the laboratory on-ramp and during her lectures surrounding the in-class workshop. Furthermore, by emphasizing "learning by doing  over and over and over again," Professor Evans explicates that she also considers the repeated use of MATLAB in the course as a resource in developing familiarity and comfort with the computational language.

In this manner, we consider Professor Evans to have designed and structured the course so that students would use computation to be a tool to conduct statistical analyses central to physics. Therefore, DC-2 connects from the design elements of \textit{Lecture}, \textit{Lab}, \textit{MATLAB references}, and \textit{Comfort w/ MATLAB from repeated practice} to the mediating process of \textit{Use MATLAB as a tool to learn physics} with positive affect.

Throughout our semester of observations, the only lecture in which Professor Evans presented code was the week before the in-class workshop, when she introduced the multi-day activity. The slides she presented covered indexed and conditional loops, both of which were explained through pseudocode and an example function. Professor Evans only spent a brief amount of time on these slides, encouraging students to refer to the course's MATLAB manual and a MATLAB help site to learn how \texttt{if}, \texttt{while}, and \texttt{for} loops worked. As was the case with all of her other lectures, this lecture predominately focused on developing students' understanding of the more advanced statistical methods used for physics analyses. Professor Evans worked through derivations that demonstrated how correlational terms surface in error terms for interrelated variables and clarified that students would need to learn more about conditional loops, directing students to MATLAB documentation and the course-specific MATLAB manual. 

In the context of this activity, we see Professor Evans situating computation as a tool to conduct statistical analyses central to physics. Therefore, DC-2 also connects from \textit{Lecture} to \textit{Extend understanding} with positive affect. In addition, DC-2 also connects from \textit{MATLAB references} to \textit{Extend understanding} with positive affect, as she outsourced the development of this computational knowledge to external resources.

\subsubsection*{Design Conjecture 3: If the utility of computation is highlighted throughout the course, students will see computation as a professional practice of physicists.}
By and large, Professor Evans didn't want students to think of themselves as "learning MATLAB;" she wanted them to view themselves as "doing experiment[s]." She repeatedly centered the practical use of computation for the field of physics, and our observations and document analysis found evidence of this framing in nearly every lecture, both in what Professor Evans said and in the datasets she used to work through sample problems. 

Specifically, during the lecture in which she introduced the multi-day activity, Professor Evans highlighted that an assumption behind students' analyses -- independence of variables -- does not always hold and used experiments from both this class and her research to demonstrate that complication. She presented the entire multi-day activity as "extending the concepts [the students have] already learned," and she deliberately had the activity build on itself, thereby highlighting the iterative nature of code development. When she later posted a Canvas announcement about the second associated homework, she reinforced the relevance of this assignment by contextualizing the capstone of this multi-day activity to students as follows: "This is our last homework and the most difficult. It takes you beyond the usual fitting and gives you an idea of the more complicated problems you will face in your future career in experimental science."

DC-3 therefore connects from the ethos of \textit{Preparation for future} and the course structures of \textit{Lecture} and \textit{Homework} to the mediating processes of \textit{View computation as professional practice of physicist} and \textit{Extend understanding} with positive affect on Figure~\ref{fig-intent}.

\subsubsection*{Process Conjecture 1: When students use MATLAB as a tool to learn physics, they develop an understanding of the use of code in physics.} 
Professor Evans believed there was a positive atmosphere surrounding to the use of computation, which she attributed to the specific intersections between deliberately scaffolded computational knowledge and statistical knowledge. Specifically, she positively assessed the way these encounters shaped students' perceptions of how code reinforced what they learned, which we consider to be cognitive PCL beliefs:
\begin{quote}
I do think that that step works as far as getting them to understand that the coding is doing what they have learned also in class and in the textbook.
\end{quote}
In what we deem an assertion of the centrality of cognitive PCL, Professor Evans drew attention to the the importance of code in developing students' understanding of physics in a computationally integrated physics course. She assessed the efficacy of these content connections in helping students come to the idea that they learn the same things through coding that they learn without code. PC-1 therefore connects from \textit{Use MATLAB as a tool to learn physics} to \textit{Performance/competence} and \textit{Cognitive beliefs} with positive affect.

In the past, Professor Evans saw that MATLAB was a differential barrier for students without computational preparation; therefore, she restructured the lab component of the course so that the first few weeks would be an on-ramp to MATLAB to help students develop comfort and familiarity with this computational language, as discussed in DC-2. She did this so students "don't have to struggle so much with the [coding] mechanics" and could instead focus on the physics. From this, she believed:
\begin{quote}
    They can start seeing the bigger picture. And if you see the bigger picture, I think you're more likely to have fun with it and to think that you're really doing experiment as opposed to how to make MATLAB work.
\end{quote}
In this passage, Professor Evans again reasserts the importance of cognitive PCL, but she does so by highlighting the importance of an underlying positive affective tone, in which students find it fun to do statistical analyses computationally. She further highlights that negative material PCL beliefs, experienced through persistent struggle with syntax, can shift students' focus away from the purpose of computation, so she designed the course structure to preclude this from happening. 
To reflect this on our conjecture map, PC-1 also connects from \textit{Use MATLAB as a tool to learn physics} to \textit{Interest} with positive affect.

\subsubsection*{Process Conjecture 2: Using MATLAB as a tool in the ways physicists use computation equips students with confidence in a relevant and transferrable skill.}
On the whole, Professor Evans did not view this as a class on MATLAB. Rather, she tried to "really put the computation as a service to the statistics" so students could benefit from gaining a transferrable skill. In what we consider to be an emphasis on cognitive PCL, Professor Evans expressed her reasoning as to why it was important for this course's central focus to be on field-specific applications of code -- students needed it for their future careers:  
\begin{quote}
You need computation to do analysis of some sort, and you want to have it be a tool that people are comfortable using so they can be comfortable using it elsewhere as well.
\end{quote}
With the intention that students would inherit positive affect from comfort with MATLAB, Professor Evans' expectations for students expanded beyond using computation in the context of this course. 

Regarding the multi-day activity, Professor Evans felt that this extension of knowledge genuinely interested students. From watching student interactions during the in-class workshop, Professor Evans believed students "liked annotating the code, not because they wanted to do a grid search, but they actually [liked] sitting with a piece of code trying to understand what it does." So, PC-2 connects from the mediating processes of \textit{Extend understanding}, \textit{Brackets around for loops}, and \textit{Annotated pseudocode} to the outcomes of \textit{Material beliefs} and \textit{Interest} with positive affect.

As noted in DC-3, Professor Evans explicitly mentioned the utility of this multi-day activity to students when she posted a Canvas announcement about the second associated homework. She contextualized the capstone of this multi-day activity by noting the relevance this new approach will have for students' experimental research careers, a stronger stance than claiming that students could or might use this advanced technique. We argue that she thereby equated the techniques used in this activity to be essential for a career in the field, as her messaging did not include hypotheticals. 
In so doing, Professor Evans communicated to students that what they were learning would be valuable outside of the course context.

Because she wanted students to feel comfortable applying code in other field-specific contexts and used the multi-day activity to help students practice that application, PC-2 also connects \textit{Use MATLAB as a tool to learn physics} and \textit{View computation as professional practice of physicist} to \textit{Cognitive beliefs} and \textit{Performance/competence} with positive affect.

\subsubsection*{Process Conjecture 3: If students connect the coding they do to the practice of a physicist, they will feel like physicists. }
In an interview, Professor Evans stated that she felt that for students, this course "is a crucial role in their development" as physicists. However, she didn't entirely see her students as `physics people,' largely because she also viewed "them as students first" because:
\begin{quote}
    There's something that I know that I can impart to them and then they can be better. And that is different than a scientific colleague.
\end{quote}
In part, this stemmed from her definition of what it meant to be a physicist, which she felt required one to be "an empiricist," because that mode of thinking was "the only way you'll get ahead in science." By designing this activity to extend students' understanding of empiricism beyond the context of the course, we consider her structural decisions to be in alignment with this desired outcome. 

Instead, Professor Evans found it more important that students saw themselves as physics people, claiming "that's what we're trying to do." Because of the changes she had made to make this course more computationally scaffolded, she reflected: "I do feel that [during the data collection] semester students felt more like that."
When we couple this with the aforementioned centrality of lab in the course, we consider this development of students' physics identity to be part and parcel of her goals. 


In an interview, Professor Evans spoke to her motivation to contextualize analytic methods through her own research, stating that she wanted students to "feel like what they're doing in the lab really is like doing an experiment and that this is what it feels like [to be a physicist]." She deliberately addressed the gap between the analyses students had used in their course and what this multi-day activity asked of them. Professor Evans connected the real-world applications to the analysis students had been doing thus far, reassuring students that "what you've done is well and good but this is the way forwards."

With her desire to contextualize the multi-day activity to help students feel like physics people, we consider the development of students' physics identities to be one of Professor Evans' goals, specifically by developing performance/competence to complete the relevant physics task of a complicated computational analysis. So, PC-3 positively connects from \textit{View computation as professional practice of physicist} and \textit{Extend understanding} to \textit{Cognitive beliefs}, \textit{Performance/competence}, and \textit{Sense of belonging}.

\subsection{Impact: Student Perspective - Bridget} \label{impact}

To investigate how Professor Evans' intentions manifested differently than intended, we study the experience of Bridget, a sophomore who intends to major in both physics and astrophysics. In a pre-course survey, Bridget reported that she came into the class knowing how to use computational tools like PhET, Wolfram Alpha, and Excel. She clarified in her initial interview that this was her first exposure to MATLAB but not her first exposure to coding, as she learned Python in a computer science (CS) course. Bridget's survey responses also indicated that prior to this semester, she had conducted research at UMN, but in interviews, she said that even though her research used code, she felt it was more for the sake of instrument control than data analysis. 

\begin{figure*}
  \includegraphics[width=\textwidth]{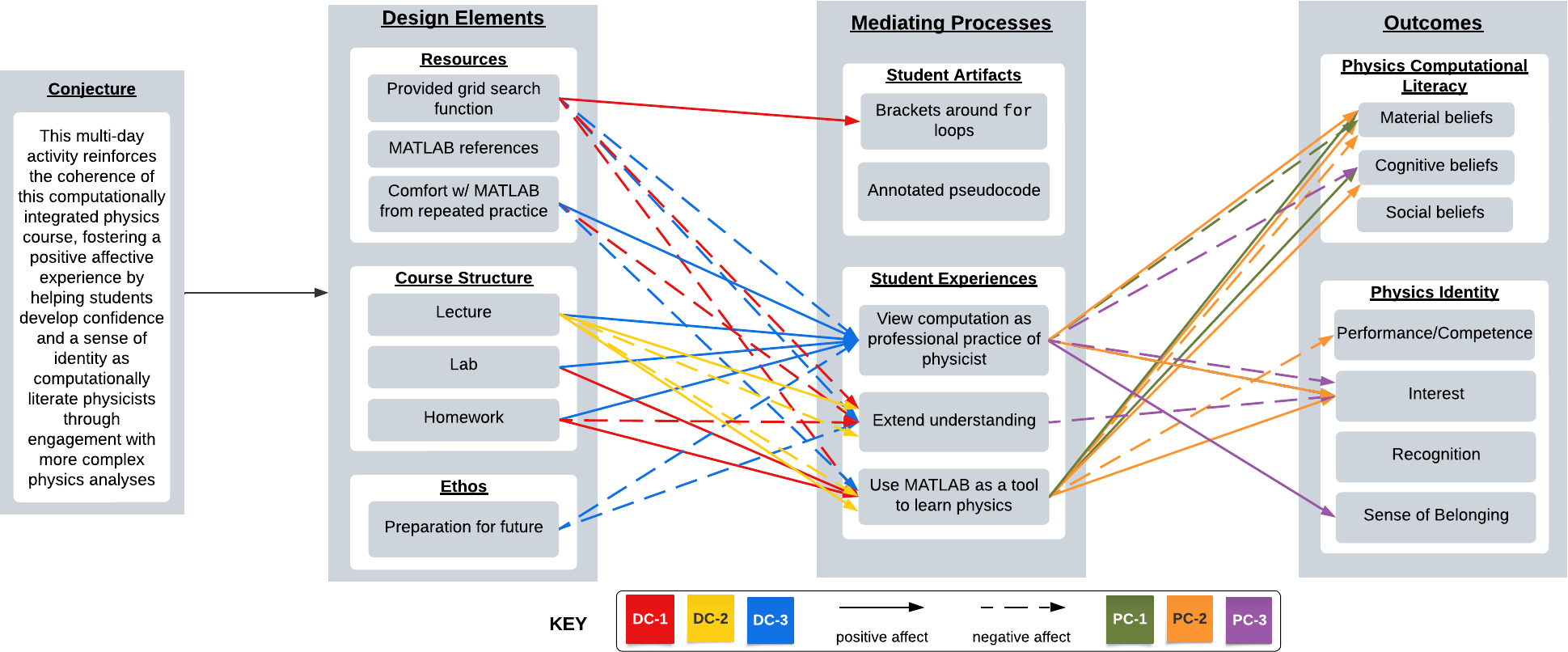}
  \caption[Bridget's conjecture map, representing impact of perceived coherence between the entire course and the activity under study]{Conjecture map representing Bridget's perspective on the multi-day activity, specifically depicting the impact of perceived coherence between the entire course and the activity under study.} 
  \label{fig-Bridget}
\end{figure*}

Our characterization of the activity's impact on Bridget's PCL and PI is given in Figure~\ref{fig-Bridget}. We use the design elements, mediating processes, and outcomes intended by the instructor (Figure~\ref{fig-intent}) so that we can analyze how Bridget's design and process conjectures are both similar and different from Professor Evans'. Again, positive and negative affect are displayed as solid and dashed lines, respectively. Specifications of data sourcing are summarized in 
Appendix \ref{Sec:Bridget_data}.

\subsubsection*{Design Conjecture 1: If computational tasks are appropriately scaffolded, then students will be able to use computation as a tool to learn physics.}
Throughout this computationally integrated physics course, Bridget noticed and appreciated how the computational tasks built upon themselves. Bridget expressed a general belief that "in this class, it's more like you need to understand the physics" rather than understand the code, so it made sense to her why "you would use [a built-in MATLAB] function instead of handwriting code to do the same thing the function does." Nonetheless, she found it helpful to begin with MATLAB's built-in linear fit function and later move to writing her own code:
\begin{quote}
MATLAB's fit function was a good introduction to simply what outputs you're looking for. So you knew that you're looking for slope because oftentimes when you do a linear fit or when you linearize your equation, you relate the slope to a group of variables. So you knew what you were looking for and then because of that, you had to code on your own, but you knew which outputs that you had to put out, so it was easier to code for those outputs.
\end{quote}
This scaffolding helped Bridget use computation as a tool to learn physics, as she felt the process of writing code developed her understanding of the physical laws she was investigating: 
\begin{quote}
Coding for it on your own made you realize exactly which variables would connect in a certain way, so you were able to see those distinctions and connections also.
\end{quote}

However, for Bridget, a large part of what made the scaffolding helpful was the way it tightly mapped to the expectations for her lab analyses. She found it "really helpful the way the homework [was] structured because that really help[ed] with the labs," especially since she could "recycle" code she wrote. So, Bridget became frustrated when she saw the relevance decrease over time as the homework moved into "smaller, tedious details that seemed a bit irrelevant when it came to what we were doing in lab and assessments," as with this multi-day activity.

To represent the tensions between Bridget's appreciation for the scaffolding's utility in the course as a whole but not in this activity, DC-1 connects from \textit{Lab} and \textit{Homework} to \textit{Use MATLAB as a tool to learn physics} with positive affect, and \textit{Homework} and \textit{Comfort w/ MATLAB from repeated practice} connect to \textit{Extend understanding} with negative affect.

Furthermore, the structural scaffolding of this multi-day activity differed from the usual sources of support Bridget came to know and appreciate. Even though Professor Evans provided the more complicated code for the 2D grid search, Bridget struggled to understand code written by someone else, as this was the first time such a task was asked of students. The first author observed Bridget bracketing the \texttt{for} loops during the in-class workshop, but she never moved beyond that level of engagement in class and did not submit the second associated homework.

To represent Bridget's engagement with the brackets, DC-1 positively connects from \textit{Provided grid search function} to \textit{Brackets around for loops}. Since Bridget's struggle with this new form of scaffolding prevented her from deeper engagement, DC-1 also connects from \textit{Provided grid search function} to \textit{Use MATLAB as a tool to learn physics} and \textit{Extend understanding} with negative affect. 

\subsubsection*{Design Conjecture 2: If students learn statistics in lecture, they can apply them when writing code.}
In general, Bridget appreciated how lecture regularly prepared her to complete her homework for this class:
\begin{quote}
I think the lecture is needed to understand the homework... like why you need that statistic that the homework is asking for. So lecture does a good job of breaking down why certain homeworks are asking you for something.
\end{quote}
Through document analysis, we determined that homework often focused on connecting the statistics underlying computational work to physical understanding by posing context-rich problems. With this in mind, we believe the centrality of statistics in lecture provided a methodological foundation that helped Bridget use computation as a tool to learn physics. 

More broadly, Bridget compartmentalized physics concepts and computational statistical analyses separately and said that she "mostly focus[ed] on the physics part." Bridget did not see computation as an iterative part of the way she came to learn physics and instead felt that "there's always a way you can figure out how to actually physically code it, but the biggest part of coding is the logic behind it. You need to know what to code for and then the actual code will come along eventually." During an interview, she expressed her stance on the use of code in physics through an analogy:
\begin{quote}
The code is your battle plan, but you need to understand what the battle is before you can initiate the plan.
\end{quote} 
She enacted this during the multi-day activity, in which she initially prioritized her own understanding of the first associated homework. Bridget reported that
"when I'm listening to lecture, I understand the concepts being told," 
 but when it came to implementing the more advanced statistics computationally, "executing it was just something I wasn't able to do." 

Reflecting Bridget's foundation of statistical knowledge from lecture, DC-2 connects from \textit{Lecture} to the mediating processes of \textit{Extend understanding} and \textit{Use MATLAB as a tool to learn physics} with mixed affect because despite her confidence from lecture, Bridget struggled to carry that forwards in the context of this multi-day activity.

\subsubsection*{Design Conjecture 3: If the utility of computation is highlighted throughout the course, students will see computation as a professional practice of physicists.}
In this course, Bridget saw how computation was used in the "career field" of physics. When Professor Evans highlighted her own research in lecture, Bridget came to realize "that these skills that you learn will help you eventually get to that point," and that she would someday be able to do Professor Evans' computational analyses. 
This was important to Bridget because even though she had taken a CS course previously, she decided not to take more CS coursework because she wanted to learn the field-specific uses rather than learning code in a CS framework, reflecting, "I don't think I would enjoy the amount of detail [in computer science] `cause I won't apply it."
To represent Bridget's appreciation for the ways in which she saw field-specific applications of code, DC-3 connects from \textit{Lecture} to \textit{View computation as professional practice of physicist} with positive affect.

Again asserting her preference for field-specific uses of computation, Bridget wanted to keep the physics at the center of this course and reported that she tried "not to rely on computation to solve the problems" she solved in homework and lab. Instead, Bridget frequently used code to validate her conceptual understanding and stated that she came to her own understanding of the physics before using computation to check her work:
\begin{quote}
The coding is a means of helping you plug and chug. And sometimes it helps you understand things. So if you change something in your code and you realize, oh, I changed that and that caused my entire physics concept to be off or something, it'll help you correlate between variables every once in a while.
\end{quote}
She became comfortable with this practice due to the success she experienced with its repeated use across course elements. Through document analysis, we found that even when asking for the use of computation, homework assignments and lab analyses could be understood with pen and paper, as Bridget preferred to do. So, DC-3 connects from \textit{Homework}, \textit{Lab}, and \textit{Comfort w/ MATLAB from repeated practice} to \textit{View computation as professional practice of physicist} with positive affect and from \textit{Comfort w/ MATLAB from repeated practice} to \textit{Use MATLAB as a tool to learn physics} with negative affect.

Nonetheless, Bridget faced challenges across the multi-day activity when she could no longer leverage her "plug-and-chug" approach to computation because she felt she struggled with syntax across the activity. To Bridget, the multi-day activity deviated from the way computation was used consistently throughout the course. This 
frustrated Bridget even though she could see the utility with regards to her professional future: 
\begin{quote}
   You can use it for non-linear fitting, but we were never asked to do that in our labs. And you could always linearize everything... Whereas it's useful information in the future, it just didn't apply to this class. 
\end{quote}
To depict Bridget's displeasure in this, DC-3 also connects from \textit{Provided grid search function} and \textit{Preparation for future} to \textit{View computation as professional practice of physicist} and \textit{Extend understanding} with negative affect.

\subsubsection*{Process Conjecture 1: When students use MATLAB as a tool to learn physics, they develop an understanding of the use of code in physics.} 
By using MATLAB in a physics context, Bridget felt that she came to understand "the interconnection between physics and coding," and she developed a general understanding of this relationship that extended beyond the scope of the course:
\begin{quote}
I can relate how physics is used in the field itself. So I have an understanding of if I go into this career, what to expect for people using programming languages with the physics aspect involved. 
\end{quote}

However, Bridget had mixed feelings on the use of MATLAB as a computational platform to achieve this goal:
\begin{quote}
You could see in your future when you're doing research, how would you use code to apply. I don't think it necessarily helped my coding resume, I would say simply because like I mentioned, MATLAB isn't a language that many people use outside of physics classes, so it helped with that understanding of how would I apply code in physics and it was a good way to introduce it, but I don't think I got a specific skill I would say.
\end{quote}
Here, Bridget attributes the use of code to its application. She can see field-specific uses of code, in the development of a positive cognitive PCL belief, but feels limited by the computational platform, a negative material PCL belief. Because she had taken a CS course before taking this physics course, Bridget developed preferences for Python over MATLAB. By the end of the course, though, she understood why MATLAB was used -- to keep the physics at the center:
\begin{quote}
It is such a heavy class already that MATLAB is a good simple language that can be used to do what we need to, it's pretty straightforward when it comes to plotting. So I think it's there for that concept issue where you need to understand how physics and programming goes hand in hand. But I feel like if we had used a more complicated language such as Python or Java or whatever it is, it would've made the class a lot more difficult than it needed to be just to get the same things done. And I think MATLAB is very intuitive when it comes to the data analysis that we're doing as opposed to Python would be.
\end{quote}
In adopting this mindset, Bridget fully embraced Professor Evans' framing of computation as a tool to learn physics and developed a positive material PCL belief. 

Considering all of these themes, PC-1 connects from \textit{Use MATLAB as a tool to learn physics} to \textit{Cognitive beliefs} and \textit{Material beliefs} with positive affect, as well as from \textit{View computation as a professional practice of physicist} to \textit{Material beliefs} with negative affect.

\subsubsection*{Process Conjecture 2: Using MATLAB as a tool in the ways physicists use computation equips students with confidence in a relevant and transferrable skill.}
While Bridget thought "MATLAB [had] been pretty easy to learn," she found that difficulties arose for her "when [she got] into details." Even though Bridget felt she understood the multi-day activity as explained in lecture, she did not understand specifically how to code the first associated homework assignment. She encountered difficulties throughout the multi-day activity, and because she was unable to resolve her syntax issues, Bridget inherited frustration across components of the activity:
\begin{quote}
It was not great for that self-confidence in physics because there are people who knew MATLAB really well and they were able to do the assignment.
\end{quote}
In other words, Bridget's confidence in her cognitive PCL contrasted with her material PCL struggles, which encouraged her to make a negative self-evaluation of her own performance/competence. We consider this to be negative affect inherited from Bridget's lack of comfort with MATLAB. Even though she was familiar with the ways they had used MATLAB in this class, her comfort did not continue in this extension activity. To represent this, PC-2 connects from \textit{Use MATLAB as a tool to learn physics} to \textit{Material beliefs} and \textit{Performance/competence} with negative affect, and it also connects to \textit{Cognitive beliefs} with positive affect.

In general, Bridget reported that she was "feeling good about now I know I need to use code and feeling good that I am understanding MATLAB," but she felt that the confidence she had developed in this course was only pertinent to the scope of this course, rather than transferrable: 
\begin{quote}
It made me feel more confident in my ability to relate coding with physics, but I also need to keep that ego in check because it's still not necessarily exactly how I would be in the career field.
\end{quote}
So, she used this activity as motivation to pursue further field-specific computational training to prepare for a career:
\begin{quote}
Something I've realized recently is that when you're going to the physics field, a lot of people just focus on the physics aspect, but more often than not, you're going to be needing code... I'm looking into coding programs or something I can do because I know I need to have a good grasp on code, especially because it's so useful in the workforce. If you can't code in the workforce, if you're looking to go in the physics or astrophysics field, then there's not much you can do.
\end{quote}
To Bridget, learning MATLAB was a step towards her future career, even if it wasn't the most applicable because compared to other languages, "MATLAB is a good transition into something that's a bit more professional." Bridget felt that this "motivates me to go on my own time to learn other coding languages like Java that are actually used." 

Because Bridget's use of computation as a tool to learn physics encouraged her to learn more about computing in the field, PC-2 also connects from \textit{Use MATLAB as a tool to learn physics} and \textit{View computation as professional practice of physicist} to both \textit{Interest} and \textit{Material beliefs} with positive affect. 


\subsubsection*{Process Conjecture 3: If students connect the coding they do to the practice of a physicist, they will feel like physicists. }
In her first interview, Bridget said that from the start of the semester, this class made her feel differently about herself than other physics classes did. Even though she quickly developed a dislike for the new way in which this class presented lab analyses, Bridget felt this clarified her career goals:
\begin{quote}
   I think it's definitely made me realize I'm not a lab person... It's helped me highlight that I don't particularly like experiments and conducting statistical analyses... So this class kind of overwhelmed me at first and I was like, I don't particularly like this, and that's okay because now I know that I don't like it. So it's helping me head into a direction.
\end{quote}

Later in the semester, Bridget justified her distaste for the way this course positioned the use of computation when she reflected, "I'm not going to be doing every physics field at once when I get into the career field." 
In so doing, we claim she found a sense of belonging by curating interest in the field through a process of elimination: determining what she didn't like.

In that same interview, Bridget reflected that in this course, computation was "helpful in understanding the basics of statistics." As she developed a belief that the connection between computing and physics fundamentally relied on statistics, she started to doubt her interest in such a future:
\begin{quote}
Because I got used to understanding the statistics that we use for these labs, it made me realize that when you're doing more research in the field, there's going to be a lot more complicated statistical analysis. So I'm still wary of that where I don't want to do that, but it's something you need to do. So I'm not necessarily looking forward to that, but I understand that you can't get away from it.
\end{quote}

PC-3 combines the outcomes of Bridget's process of self-discovery through process of elimination by connecting from \textit{View computation as a professional practice of a physicist} to \textit{Sense of belonging} with positive affect and to \textit{Cognitive beliefs} and \textit{Interest} with negative affect. 

During the end-of-semester interview, Bridget simultaneously held these tensions with her own understanding of the purpose of the multi-day activity. 
Even though she knew this multi-day activity was designed to provide a more realistic glimpse into the career of a physicist, it didn't connect to the tightly coherent ways in which she had used linear fitting in this course, so she thought that challenge "added a bit more stress than it needed to" and was "a bit unnecessary" in the context of this course. To Bridget, the multi-day activity felt like a tack-on, rather than a culmination of a holistic computationally integrated physics course.

To represent Bridget's frustration towards the activity's extension beyond the analysis methods she typically used in lab, PC-3 also connects from \textit{Extend understanding} to \textit{Interest} with negative affect.

\subsection{Cross-case Comparisons}
Having described and analyzed each individual case, we now move to a cross-case analysis~\cite{alma9975672698201701} to compare Professor Evans' intentions for students' affective experiences with Bridget's own experience. To do so, we re-present the manifestations of each set of design and process conjectures by summarizing our analysis of each perspective from Sections \ref{intent} and \ref{impact}, rather than presenting raw data. 

\subsubsection*{Design Conjecture 1: If computational tasks are appropriately scaffolded, then students will be able to use computation as a tool to learn physics.}
Professor Evans deliberately scaffolded how students learned computational techniques across lab and homework, and Bridget capitalized on that, developing a belief that in physics, the creation of code was deprioritized relative to the use of code as a tool. In this capacity, we see alignment of intent and impact regarding the foundation on which the activity was to build, as the structural reinforcement of scaffolding across the course contributed positive affect towards Bridget's perception of MATLAB as a tool to learn physics.

However, Bridget became reliant on the course's scaffolding and struggled across this multi-day activity, where the forms of support differed from how they had consistently been used in the course. While Professor Evans thought students enjoyed the annotation process, Bridget became frustrated as she struggled to understand code written by someone else. Bridget did engage with the in-class workshop by bracketing the provided grid search function, as Professor Evans intended, but doing so did not shape Bridget's self-perception. The support that the provided function offered did not extend Bridget's understanding beyond typical course content and instead left her feeling disheartened and confused. Professor Evans thought that the provided grid search function would provide a more positive form of scaffolding during the multi-day activity than Bridget experienced, but this activity-specific form of the conjecture was not realized.

Even though this conjecture had an alignment between intent and impact for the course as a whole, that coherence did not extend to the multi-day activity. As such, DC-1 highlights the challenges and affordances of incorporating activities that push beyond the course's established methods and frameworks.

\subsubsection*{Design Conjecture 2: If students learn statistics in lecture, they can apply them when writing code.}
Integrating computation into an existing course with content and design made Professor Evans feel like she needed to keep lecture focused on the physics. So, she prioritized cognitive PCL in her instruction and 
outsourced support with material PCL to MATLAB references. This divide was what contributed negative affect towards the way Bridget chose to engage with the multi-day activity, even though it provided positive affect relative to the course as a whole. 

Bridget largely considered code as a separate entity from her conceptual knowledge of physics and felt positively about doing so. Despite her understanding of the concepts, Bridget still struggled to execute them computationally. This insurmountable hurdle, contrasted with lecture's focus on concepts, provided Bridget with mixed affect towards the use of MATLAB as a tool to learn physics and towards the extension of her own understanding.

Notably, DC-2 was fulfilled through some anticipated design elements, but Bridget did not use all of the resources Professor Evans intended to serve as additional support and struggled more than her professor anticipated. This draws attention to the challenges of tying scaffolding to broader frameworks when integrating computation into a course.

\subsubsection*{Design Conjecture 3: If the utility of computation is highlighted throughout the course, students will see computation as a professional practice of physicists.}
To enact this conjecture, Professor Evans highlighted her own research in lecture so that students could view themselves as doing real physics experiments. For Bridget, this revealed the gap between her skillset and Professor Evans', but Bridget knew that, across course elements, she was building the skills necessary to bridge this gap. 

Bridget generally appreciated the emphasis on the field-specific uses of code, but the way in which she regularly used code throughout the course differed from the ways Professor Evans framed it. Bridget reported using code to "plug and chug" after coming to her own understanding of the physics, whereas Professor Evans described it as an integral part of the analysis process and demonstrated the iterative nature of code development and similarly structured the project to build upon itself. When the multi-day activity aligned more with Professor Evans' understanding of computation than Bridget's, 
Bridget felt a deviation from the consistent use of computation. Professor Evans' overt messaging emphasized how this assignment extended what students had already learned, which counteracted but did not overcome Bridget's expectations for continuity, manifesting in unintended negative affective connections. 

As with DC-1, DC-3 demonstrates the potential for unexpected negative affect when using activities that students perceive to extend beyond course content.

\subsubsection*{Process Conjecture 1: When students use MATLAB as a tool to learn physics, they develop an understanding of the use of code in physics.} 
Centering cognitive PCL, Professor Evans wanted students to be able to use MATLAB as a tool to focus on the physics and make a positive self-evaluation of their performance/competence in doing so. By both understanding and appreciating the use of code in physics, Bridget developed a positive cognitive PCL belief. 

Professor Evans also felt it was important for students to not struggle with the computational platform, which Bridget noted positively in highlighting how MATLAB allowed her to focus on the physics. The course's emphasis on the ways in which physicists use code posed a disconnect for Bridget, as she formed a negative material PCL belief in feeling that the choice of computational platform, MATLAB, was not the most transferrable post-college. This created mixed affect as Bridget experienced both positive and negative material PCL beliefs when the course messaging addressed the use of code in general but did not emphasize the utility of the specific coding language.

As such, PC-1 highlights the ramifications of instructional design decisions, like the choice of computational platform, on students' perceptions of a computationally integrated physics course.

\subsubsection*{Process Conjecture 2: Using MATLAB as a tool in the ways physicists use computation equips students with confidence in a relevant and transferrable skill.}
Professor Evans wanted students to leverage their comfort with MATLAB to gain confidence in an analysis that professional physicists would conduct. However, Bridget's comfort with MATLAB did not extend beyond how she was regularly used code throughout the course. Bridget developed confidence from a consistent, repeated use of MATLAB as a tool and said that she only struggled with the MATLAB-specific details, which she felt were the focus of the multi-day activity. Consequently, Bridget couldn't code the concepts she felt she understood. So, she developed negative affect towards her own performance/competence through the development of a negative material PCL belief, despite her positive cognitive PCL belief.

Specifically, Bridget felt that the use of MATLAB as the course's computational platform limited the transferability of this experience. Rather than demotivating her, this served to inspire Bridget to learn other computational languages that would tie in better to field-specific uses of computation, bolstering her interest. As she was able to see the connection to the ways in which physicists would use code, Bridget simultaneously held a positive material PCL belief and appreciated the way this course helped her develop.

In specific, PC-2 calls attention to the potential for the perceived coherence of a computationally integrated physics course to have affective ramifications.

\subsubsection*{Process Conjecture 3: If students connect the coding they do to the practice of a physicist, they will feel like physicists. }
Professor Evans believed that this multi-day activity would connect computing to the professional practices of physicists by extending students' understanding, which would in turn help students see themselves as `physics people' through performance/competence in cognitive PCL. She felt that with this connection, students would be able to  develop a sense of belonging in the field.

Bridget saw Professor Evans' contextualization of the activity, and she agreed that the skills she would gain from this activity would be relevant in the future. However, she struggled to see the utility of this activity in the context of this course, as it differed from the ways they had regularly used code. 

As Professor Evans wanted to convey, Bridget learned that statistics were an inherent part of physics analyses, but because of Bridget's deep familiarity with the analysis methods she used in lab, she felt unnecessarily stressed by knowing that the real world would be more complicated than what she was asked to do. Because Bridget was not able to use this multi-day activity to extend her understanding as Professor Evans intended, Bridget saw the gap between what she was doing in this class and what she would do in her future as a physicist and negatively evaluated her cognitive PCL and interest in the field. 

In part, this surfaced because throughout the class, Bridget gained clarity on what she wanted to do in her future, but she did so by determining what she did not enjoy in this class. So, she developed negative affect towards her own interest but felt a sense of belonging by narrowing down her options.  

The misalignment that surfaced in PC-3 highlights the gap that exists between students and apprentice researchers and the challenges in bridging it with positive affect.
\section{Discussion}\label{Sec:discussion}
While Professor Evans deliberately structured this multi-day activity to be an affirming encounter with real-world physics, Bridget struggled to see the activity as coherent with the course as a whole because she faced difficulties that impeded her ability to see herself as a physicist. 

For any course to achieve deep coherence, situating a task in a relevant context alone is inadequate for high levels of meaningful student engagement~\cite{aikenhead2003review}. The concept of relevance is multi-dimensional and varies for individual learners~\cite{stuckey}. 
In a computationally integrated introductory physics course for non-majors, Nair and Sawtelle conceptualized relevance as co-constructed by students and their environments~\cite{nair2019operationalizing}, and our case study demonstrated that this also occurs in a computationally integrated physics course beyond the introductory level. The environment in which Bridget engaged throughout the semester regularly rewarded her for certain habits of behavior and epistemological stances on the use of code, and she carried forwards her `plug and chug' conceptualization of code, despite the additional challenge this activity posed. Further, when this activity deviated from the course's tight coherence, Bridget viewed it as more relevant for her future than for her present. As such, we claim that the perception of relevance, reflected by interest, was also shaped by Bridget's prior engagement and reinforced by earlier success. 

A regression study found that the interest of middle- and high-school students in physics as a school subject was more related to their self-esteem as high achievers than to their interest in physics as a subject~\cite{haussler2000curricular}. 
Along those lines, our case study also nuances the findings of Kinnuen and Simon, who highlighted the importance of building self-efficacy early in a programming course~\cite{kinnunen2011cs}. By analyzing an activity at the end of a course, we found that the affective foundation upon which Bridget built did not ensure an extension of the same affective experience and was instead detrimental to her self-perception in this multi-day activity. 

In computationally integrated physics courses, students who know how to code still struggle~\cite{hamerski}, so the question becomes how can we best support them? If we take an affective approach, we note that prior literature on introductory-level computationally integrated physics courses has found that students did not consider themselves to be coding and to instead be doing physics~\cite{bumler2019previous}. This may well mark the success of an integrated curriculum, but it also questions how code is leveraged. In this study, we found that Bridget did view herself as coding and was simultaneously able to focus on the physics -- until she hit a challenge. In studies of more introductory computationally integrated physics courses, debugging has been found to be a common student difficulty~\cite{caballero2014integrating}, and despite Bridget's prior computational training,  this was also the case for her.  Caballero, et al. attributed this difficulty to a "somewhat loose integration of computational modeling in their physics course," and in this context, we believe that the tightness of the integration has two levels: the course and the activity. Even though computation was tightly integrated throughout the activity, it was integrated in a manner that diverged from the ways in which computation was integrated in the course as a whole. 

Similarly, when offered a new source of support, like the provided grid search function, Bridget did not take advantage of it because she was not equipped to. This was the first time this course asked her to use code written differently than she would have written, and Bridget struggled to understand how the code worked, even though the provided function was intended to alleviate that issue. Other studies have found that scaffolding can either remove the difficulty of an exercise not removed rapidly enough or be beneficial if introduced early~\cite{fennell2019designing, vieira2015exploring}. The implementation of this scaffolding matters immensely, as one study attributed students' difficulties with developing understanding late in a course to the rate at which computational projects became more advanced, and we find resonance with this in Bridget's experience~\cite{vieira2015exploring}. Professor Evans attempted to address a gap in students' computational knowledge, but we consider this to fall into what Vieira, et al. refer to as "providing too much computational support." Their interest was in student understanding and performance, but our case study demonstrated similar outcomes with affect.

Professor Evans deliberately enacted her intentions for this activity so that it would be a culmination of the course in which students would act like and feel like physicists. The way in which physics content is contextualized in a course has been found to play a role in students' affective experiences. H{\"a}ussler and Hoffmann found it helpful for students to connect what they were learning to situations they would encounter outside of class~\cite{haussler2000curricular}. Our case study nuances this, because as students move through their physics coursework, they begin to see a more clear professional future. For Bridget, this meant that she felt she would not need to use everything she learned in class, even though, in alignment with existing research~\cite{caballero_computation_2015, zwickl2017characterizing}, Professor Evans considered these computational skills to be useful for a future in physics.

As such, this case study draws attention to the understandable but challenging gap between being a physicist and being a student. 
What does this distance do for students, demystify or inspire? For Bridget, it was some of both, and conjecture mapping helped to reveal how and where she developed a negative affective stance. 
Curating a positive affective experience is key to curating meaningful engagement for students~\cite{jaber2016learning}. Situating our case study in a computationally integrated physics course, we saw the emergence of a well-known pattern of engagement from traditional physics classrooms when Bridget referred to her use of code as "plug and chug." This mode of student engagement has shown up in physics education research studies of coherent activities (e.g.~\cite{tuminaro2007elements}), frustrating both researchers and practitioners alike for reasons that have more to do with epistemological framing than 
lack of productivity, as they can carry forwards throughout a student's education.

Still, we must keep in mind that this occurred in a coursework structure. After all, Bridget is a student, not a professional physicist, and she was behaving as a student. 
Similarly, Professor Evans hoped her students would go above and beyond expectations for baseline engagement, as most professors would. 
This case study demonstrated how specific computational design decisions can keep students in novice rather than journeyman epistemological frames, tying into the work of Bing and Redish~\cite{bing2012epistemic}. Further, the overarching affective tone can shape the extent to which students see themselves as "doing science" rather than "doing school"~\cite{jimenez2000doing}, but Bridget's experience demonstrated that a solid foundation in this framing throughout a course does not guarantee the application in a new activity. 
Instead, we claim that she could not adjust her framing to encompass an engagement with code that she did not write herself, representing a gap between her practical epistemology, in which she developed beliefs about her own work, and her formal epistemology, in which she developed beliefs about the field of physics~\cite{sandoval2005understanding}. When facing a challenge, Bridget's epistemological framing kept her in the role of a student, where she attempted to experience success through comfortable patterns of behavior. 
Why did the grid search function not expand her epistemological framing of the activity? We investigate this in our companion paper, where we analyze the experiences of Bridget and a classmate who had a differing affective experience with the activity~\cite{mchale2026narrative}.

Even though this case study focused on a multi-day activity in a computationally integrated physics course, the structure of the activity resembled a common use of computation. Most physics instructors that use computation do so in either isolated homework assignments or projects, both of which place the burden of computational knowledge outside class time~\cite{caballeromerner}. Since the present analysis focuses on such an isolated activity, it may be well suited to those instructional settings, and may be well suited to furthering understanding of these educational contexts. 
\section{Limitations}\label{Sec:limitations}
Our goal in this paper was to develop a rich understanding of a particular manifestation of an activity within a computationally integrated physics course, not to generalize across contexts. We therefore deliberately situated our study to provide sufficient depth to both capture and map the intricacies of affective engagement with an activity. We do not claim that all students engaging with an extension activity at the end of a computationally integrated physics course would have the same affective experience as Bridget, nor do we claim that all computationally integrated physics courses create the same foundation upon which an extension activity builds. We acknowledge that as a case study, this work is inherently limited in its generalizability, but we hope our rich contextualization has produced the resonance necessary for readers to determine the transferability of our findings to their own settings~\cite{tracy2010qualitative}. 

In addition, the use of conjecture mapping imposed multiple limitations on our findings. Specifically, we traced two perspectives on the same conjecture. Our decision to investigate the affective outcomes stemming from this activity's connection to the computationally integrated course in which it was situated was driven by the data and was a result of analytic memoing. In this spirit, we claim our study investigates a phenomenon intertwined in the data and remains close to the data throughout. There are other insights into the data that would motivate the selection of different initial conjectures and affective outcomes, but we chose to present a narrow scope through a small number of perspectives to clearly articulate the creation of our conjecture maps.

\section{Future Work}\label{Sec:future_work}
In this paper, we characterized impact through the experience of one student, and in our companion paper, we complicate our definition of impact through multiple student experiences with this same activity~\cite{mchale2026narrative}. In particular, the ways in which this analysis highlighted the role of negative affect from the use of resources, engagement with structure, and an overarching ethos motivates us to investigate how students' framing of provided code shapes the development of affective outcomes.

This work also operationalized intent through the perspective of an individual professor. However, the professor was not the only instructor of this course, as graduate student TAs played critical roles in students' experiences in the connected lab sections. In the future, we plan to study the role of such TAs in computationally integrated courses, both with and without associated labs.

In a study of a computationally integrated physics curriculum, Serbanescu, Kushner, and Stanley found that students benefited the most from computation in subsequent courses, when they could leverage the computational techniques they learned~\cite{serbanescu2011putting}. As this paper is part of preliminary investigations into a broader departmental change, we also plan to study how students experience multiple computationally integrated courses individually, concurrently, and in sequence.

To secure an equitable future for the field of physics, we need to ensure our teaching modalities positively impact students' self-perception. Broadly, we hope to see more PER studies investigating the efficacy of curricular design by centering affect in student experience, and we look forward to such scholarship that specifically focuses on intersections between computing and physics.

\section{Conclusion}\label{Sec:conclusion}
Intent and impact are different, and it is important to capture both when studying or evaluating a learning environment. Conjecture mapping revealed how the curricular structure played a role in Bridget's engagement with the scaffolded activities and movement towards affective outcomes. Professor Evans' deliberate design of this multi-day computational activity did not entirely have the results she anticipated. The curricular structure played a role in Bridget's framing of the scaffolded activities, as she wasn't able to learn from her mistakes. For Bridget, the consistency with which computing was reinforced across course elements differed in this activity, compared to the rest of the semester. As a result, she struggled to see the utility of this activity in the context of this course, which contributed to her negative self-perception.

Of specific note, the professional relevance of this activity did not outweigh the insufficient scaffolding. There were many computational goals in this activity, but they worked against each other and negatively shaped Bridget's self-perception. She inherited unresolved issues that were not addressed as the activity built on itself, and she lacked sufficient time to learn from her mistakes, which negatively impacted her self-perception.

For researchers, this case study nuances previous findings about computational multilinguality. As found in other studies, we find that prior exposure to computation impacts student resilience~\cite{LaneHeadley}, but analyzing them through PCL separated the difficulties and affordances of prior computational knowledge. Bridget's beliefs about computational literacy stemmed from material skills with one language, but these did not allow her to transcend material challenges with the course's computational platform. Even though Bridget could recognize that a robust PCL is language agnostic, her epistemological framing did not expand. 
This raises further research questions about positioning computation as a professional practice in physics courses. For Bridget, too much of a taste of the future wasn't helpful. What impacts the point of diminishing returns for perceived utility? 

For practitioners, this case study highlights how the coherence of a computationally integrated course shapes the perceived utility of included activities. While Bridget knew that the activity was relevant for a professional career in physics, she didn't see it as relevant in the context of this course because it did not directly map to the ways in which she developed code throughout the course. This opens a new line of questioning: What is the optimal extent of a professional practice? Would it have been better to reign in the computational goals of the course? Studying Bridget's path towards affective outcomes highlights the power in understanding why individual students struggle so we can revise material in ways that more accurately meet intended outcomes. Professor Evans was well-versed in this, and this case study demonstrates the reality that the work of teaching is never done. There are always ways to help more students benefit from alternative pathways to success, but to create those, we need to understand the ways in which instructional intent and student impact differ.  

\acknowledgments{}
This material is based upon work supported by the National Science Foundation under Grant No. 2236244 and 2237827. Any opinions, findings, and conclusions or recommendations expressed in this material are those of the author(s) and do not necessarily reflect the views of the National Science Foundation.

In addition, we acknowledge funding from the University of Minnesota (UMN) Norwegian Centennial Chair Travel Fellowship, which enabled an extended research visit at the Center for Computing in Science Education (CCSE) at the University of Oslo. We thank members of the CCSE for theory-building conversations, and Associate Professor Scharber of UMN's Department of Curriculum and Instruction for generative discussions about methodology.

\FloatBarrier

\bibliography{bibliography.bib} 





\appendix

\section{Data Sources for Professor Evans' Conjecture Map}
\label{Sec:Evans_data}
\begin{center}


\begin{tabular}{|c|c|c|c|c|c|c|}
\hline
\textbf{} & \textbf{DC-1} & \textbf{DC-2} & \textbf{DC-3} & \textbf{PC-1} & \textbf{PC-2} & \textbf{PC-3}\\
\hline
\makecell{\textbf{Early Semester} \\ \textbf{Interview}} &   & x &    &   & x &    \\
\hline
\makecell{\textbf{End-of-Semester} \\ \textbf{Interview}} & x &   & x & x & x & x \\
\hline
\textbf{Observation} & x & x & x  &   &   & x  \\
\hline
\makecell{\textbf{Document} \\ \textbf{Analysis}} & x & x & x  &   &   &    \\
\hline
\end{tabular}
\captionof{table}{Data sources used for our representation of Professor Evans' design and process conjectures.}
\label{tab:prof}
\end{center}

\section{Data Sources for Bridget's Conjecture Map}\label{Sec:Bridget_data}
\begin{center}
\begin{tabular}{|c|c|c|c|c|c|c|}
\hline
\textbf{} & \textbf{DC-1} & \textbf{DC-2} & \textbf{DC-3} & \textbf{PC-1} & \textbf{PC-2} & \textbf{PC-3} \\
\hline
\hline
\makecell{\textbf{Early Semester} \\ \textbf{Interview}} & x & x & x &   & x & x  \\ \hline
\makecell{\textbf{End-of-Semester} \\ \textbf{Interview}} & x & x & x & x & x & x \\
\hline
\textbf{Observation} & x &   &   &   &   &    \\
\hline
\makecell{\textbf{Document} \\ \textbf{Analysis}} & x & x & x  &   &   &    \\
\hline
\end{tabular}

\captionof{table}{Data sources used for our representation of Bridget's design and process conjectures.}
\label{tab:student}
\end{center}
\end{document}